\documentclass[mathleft
]{an}
\usepackage{graphicx}
\usepackage{times}
\begin{document}

\Pagespan{362}{}
\Yearpublication{2007}%
\Month{3}%
\Volume{328}%
\Issue{3-4}%
\DOI{10.1002/asna.200610743}%

\title{The Solar Orbiter mission and its prospects for helioseismology}

\author{J. Woch\inst{1}\fnmsep\thanks{Corresponding author:
  \email{woch@mps.mpg.de}\newline}
\and  L. Gizon\inst{1}
}
\titlerunning{The Solar Orbiter Mission}
\authorrunning{Woch \& Gizon}
\institute{
Max-Planck-Institut f\"ur Sonnensystemforschung, Max-Planck-Str. 2,
D-37191 Katlenburg-Lindau, Germany
}

\received{22 Dec 2006}
\accepted{2 Jan 2007}
\publonline{later}


\abstract{
Solar Orbiter is intended to become ESA's next solar mission in heritage of
the successful SOHO project. The scientific objectives of the mission, its
design, and its scientific payload are reviewed. Specific emphasis is
given to the perspectives of Solar Orbiter with respect to helioseismology.}

\maketitle

\section{Introduction}
The successful ESA/NASA solar observatory SOHO which was launched in 1995
has provided and still is providing an
unprecedented breadth and depth of information about the Sun,
from the Sun's interior, through the hot and dynamic atmosphere,
to the solar wind and its interaction with the interstellar medium.
The SOHO mission still constitutes the prime focus for the research of the
European solar science community.
Specifically, data returned from the MDI instrument onboard SOHO
have given a tremendous push to the field of
helioseismology.
The Solar Orbiter mission has been configured as the logical
advancing step in the exploration of the Sun after SOHO.
By approaching the Sun as close as 48 solar radii ($0.22$~AU),
Solar Orbiter will view the solar atmosphere
with a spatial resolution of less than 150 km (1 arcsec) and will perform
the closest ever in-situ measurements.
In the course of the mission the inclination of the spacecraft orbit
to the ecliptic will incrementally increase to reach heliographic latitudes
larger than 30$^\circ$ which will provide the first
ever views of the Sun's polar regions.
A further unique aspect of the mission is the combination of comprehensive
remote sensing and in-situ instrumentations allowing to trace solar features
from the photosphere to interplanetary space.

Of high interest for helioseismology is the large range of
spacecraft-Sun-Earth angles covered by Solar Orbiter. In combination with
observation from Earth (ground-based or near-Earth orbit), Solar Orbiter will
offer the opportunity for stereoscopic helioseismology.

\section{Scientific objectives}
Missions such as Helios, Ulysses, Yohkoh, SOHO, TRACE and RHESSI have advanced
significantly our understanding of the various atmospheric layers of the Sun,
the solar wind and the three-dimensional heliosphere.
Further progress is to be expected from missions like STEREO, Solar-B,
and the Solar Dynamics Observatory (SDO), the first of
NASA's Living With a Star (LWS) missions.

Each of these missions has a specific focus, each fulfilling its part
within a coordinated solar and heliospheric research effort.
Solar Orbiter with its unique orbit and advanced suit of in-situ and remote
sensing instruments constitute an important augmentation to this effort.
Solar Orbiter will for the first time:\\
- Explore the uncharted innermost regions of our solar system\\
- Study the Sun from close-up\\
- Fly by the Sun tuned to its rotation and examine the solar surface
and the space above from a corotating vantage point\\
- Provide images and spectral observations of the Sun's polar regions
from out of the ecliptic.

According to the ESA Solar Orbiter Science Requirement Document the top-level
scientific goals of the Solar Orbiter mission are to:\\
- Determine the properties, dynamics and interactions of plasma,
fields and particles in the near-Sun heliosphere\\
- Investigate the links between the solar surface, corona and inner
heliosphere\\
- Explore, at all latitudes, the energetics, dynamics and
fine-scale structure of the Sun's magnetized atmosphere\\
- Probe the solar dynamo by observing the Sun's high-lati-tude
field, flows and seismic waves.

\section{Mission design}
Solar Orbiter is a three-axis stabilized spacecraft.
On the sun-facing side a heat shield protects the spacecraft and its payload
from the compared to Earth 25 times higher radiation. The heat shield has to
provide the apertures for the remote sensing instruments.

According to the current baseline (as at December 2006) Solar Orbiter is
scheduled to be launched in May 2015 on a Soyuz-Fregat 2-1b.
The spacecraft will use a chemical propulsion system and multiple
gravity assist manoeuvres at Venus and Earth to reach its science orbit.
The initial cruise to the science orbit will last 3.4 years.
In the science phase, Solar Orbiter will be in a
3:2 resonant orbit with Venus (period 149.8 days). The Venus fly-bys are
used to slowly increase the inclination of the orbit.
The total mission duration will be approximately 10 years.
The minimum perihelion distance during the science phase will be 48 solar radii
($0.22$~AU) with a maximum solar latitude of 34{$^\circ$}.
The projection of the baselined orbit into the ecliptic x-y plane and the solar
latitude as a function of mission duration are shown in Figure 1
and Figure 2, respectively.

\begin{figure}
\includegraphics[width=\linewidth]{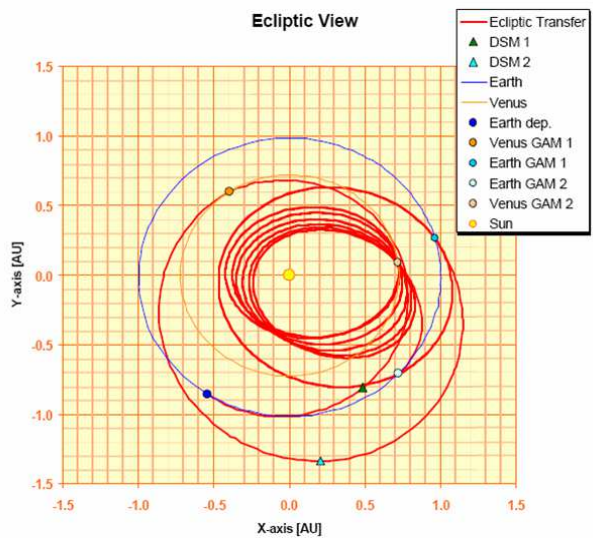}
\caption{The projection of the orbit of Solar Orbiter into the ecliptic plane.
Earth and Venus orbit as well as the position of various spacecraft
manoeuvres are indicated. (Figure adapted from a presentation by R. Marsden.)}
\label{label1}
\end{figure}

\begin{figure}
\includegraphics[width=\linewidth]{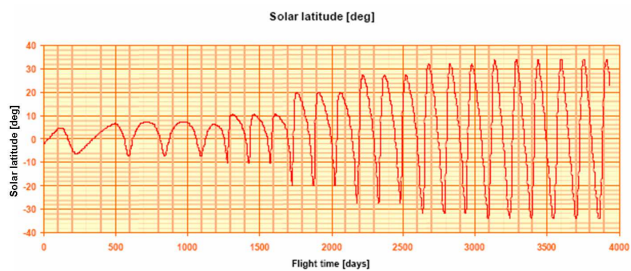}
\caption{The solar latitude of Solar Orbiter as a function of mission duration.
(Figure adapted from a presentation by R. Marsden.)}
\label{label2}
\end{figure}

Figure 3 shows the time profile of the maximum solar latitude reached in
each orbit together with a predicted profile of the solar activity based
on the sunspot number.
In 2015, Solar Orbiter will be launched into the solar activity
minimum, with the cruise phase extending through the minimum phase. Solar
activity is expected to increase again at the start of the science phase.
While the activity further develops, Solar Orbiter will advance to higher
latitudes. During the declining phase of the activity cycle the spacecraft
will reach highest latitudes. This offers the possibility to study the
transition from an active to a quiet Sun from a high-latitude vantage point.
The current mission baseline allows observations to be carried out for about
10 days around the perihelion passes and/or the high-latitude segments
of each of the science orbits.

\begin{figure*}
\includegraphics[width=130mm]{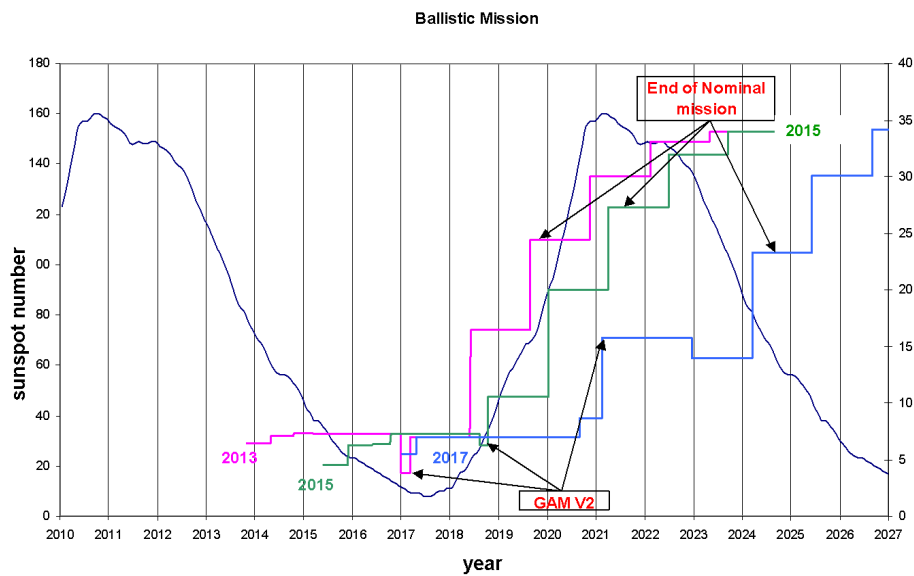}
\caption{The maximum solar latitude reached for each of the science orbits as a
function of mission duration for various launch
dates together with the predicted sunspot number.
(Figure adapted from a presentation by R. Marsden.)}
\label{label3}
\end{figure*}

\section{Scientific payload}
The current Solar Orbiter scientific payload ia a comprehensive set of remote
sensing and in-situ instrumentation. It offers the unprecedented
opportunity for complementary observation of solar processes as they evolve
from the photosphere into near-Sun interplanetary space.
The near-Sun interplanetary measurements together with simultaneous
remote sensing observations of the Sun will permit to disentangle spatial
and temporal variations during the corotation phases. They will
allow to understand the characteristics of the solar wind and
energetic particles and link them to the plasma, neutral atmosphere
and magnetic conditions in their source regions on the Sun.
The Solar Orbiter remote sensing instruments will deliver images of the
solar atmosphere with a spatial resolution of $0.5$~arcsec pixels,
equivalent to ~80 km on the Sun at $0.222$~AU,
which means resolving solar features as small as 160 km
(ESA Solar Orbiter Science Requirements Document).
The reference core payload, their scientific goals and allocated resources
(according to ESA Solar Orbiter Payload Definition Document) are shown in Figure 4.

\begin{figure*}
\includegraphics[width=170mm]{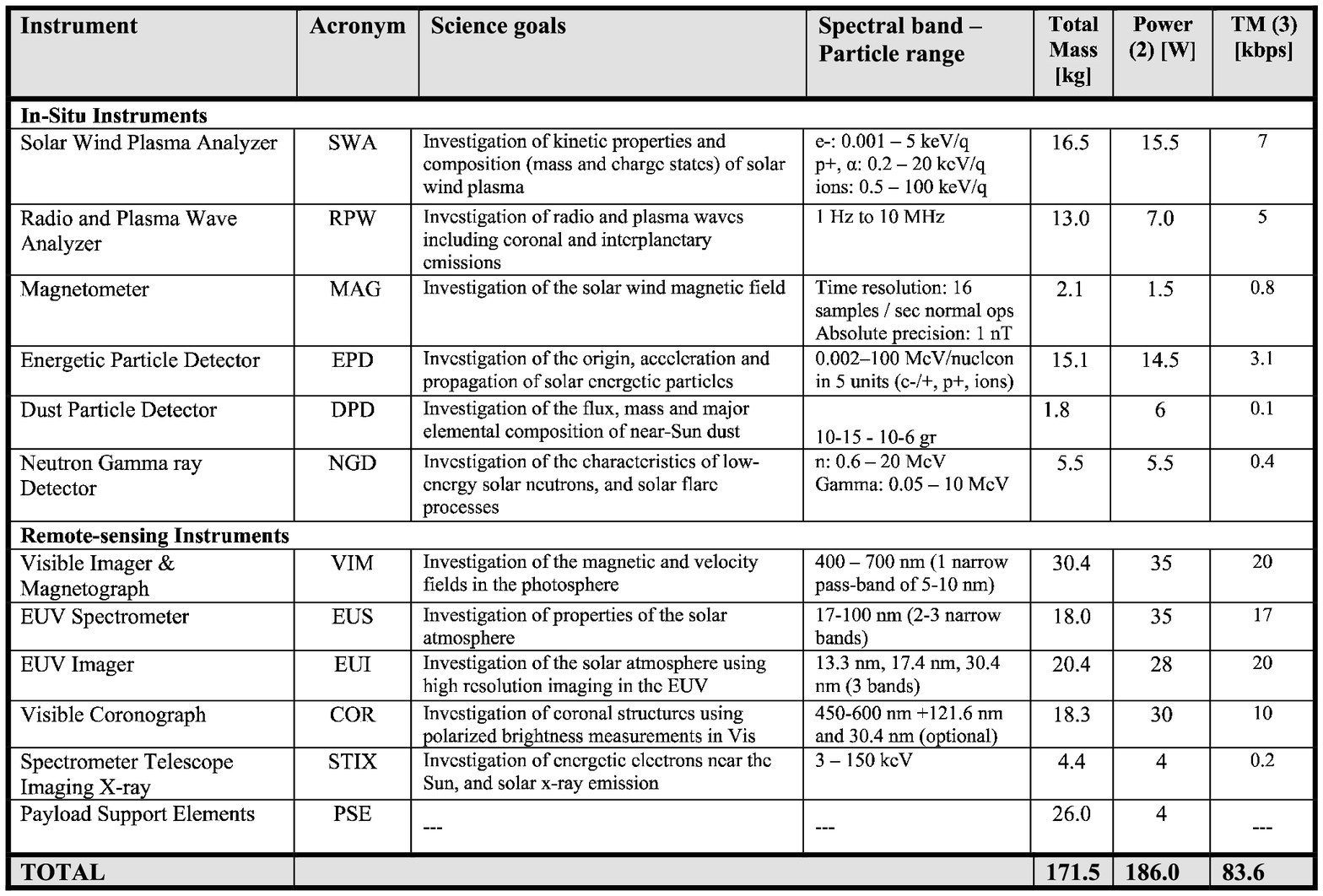}
\caption{The scientific core reference payload of Solar Orbiter
(adapted from the ESA Solar Orbiter Payload Definition Document).}
\label{label4}
\end{figure*}

\section{Current status}
Solar Orbiter is currently targeted for a launch in 2015. In July 2006 ESA
has issued a letter of intent (LOI) to propose instruments for the Solar Orbiter
mission with deadline for responses in September 2006. ESA has received 23
LOIs, documenting the interest of the solar community in Solar Orbiter.
However, due to the financial constrains of ESA, the mission is currently
reviewed again and
undergoes a consolidation phase with the aim of cost reduction.
Beside descoping, one potential way of limit costs to ESA is to
merge Solar Orbiter with the Sentinel mission of NASA.
The consolidation phase will be used to advance the readiness of the mission,
specifically in respect to the critical heat shield - instrument interfaces
by further refinement of the instruments based on interactions
with potential principal investigator teams.
Final decision on Solar Orbiter is expected in November 2007.

\section{The VIM instrument and its perspective for helioseismology}
The Visible Light Imager and Magnetograph (VIM) is one of the remote sensing
instruments in the core reference payload of Solar Orbiter.
VIM will measure the structure, dynamics and evolution of the full
magnetic vector and of the flow field in the solar
photosphere as well as waves and
oscillations that penetrate into the solar interior.
VIM will give the first view of the
magnetic and velocity field in the polar regions.
VIM consists of two telescopes, a High Resolution Telescope (HRT)
and a Full Disk Telescope (FDT), both of which feed a filtergraph.
They provide high resolution measurements
within a limited field-of-view and low resolution
measurements covering the full solar disk.
Maps of the the full photospheric magnetic field vector,
the line-of-sight velocity, and the continuum intensity will be obtained
with a cadence of less than a minute.
This requires recording linearly and circularly polarized
intensities (Stokes vectors) at different
wavelengths in a Zeeman-sensitive photospheric
spectral line and the nearby continuum.

Among the Solar Orbiter instrumentation, VIM will deliver the most
relevant data for helioseismology.

The method of global-mode helioseismology (Christen-sen-Dalsgaard 2002)
has allowed to map the solar differential rotation as a function
of latitude and radial distance throughout most
of the convection zone with a very good level of precision,
while the sound speed profile averaged over spheres
has been measured down to the solar core. At high
heliographic latitudes, however, the picture of
the solar interior is far from complete
and is fuzzier with greater depth. In
particular, global mode frequency inversions for rotation
cease to be accurate above 70$^\circ$ latitude and are uncertain
in the radiative interior.

The developing science of local helioseismology (Gizon \& Birch 2005),
which aims at producing three-dimensional images of
the solar interior, has provided information in the upper
third of the convective envelope and only at
latitudes less than 50$^\circ$. High latitudes are
difficult to study because the sensitivity of the line-of-sight
Doppler measurements to the nearly radial pulsations
drops as a function of center-to-limb distance. In
addition, the reduction of the effective spatial sampling
on the Sun toward the solar limb implies that moderate to
high degree modes are more difficult to detect away from
disk center in the center-to-limb direction.
That measurements are difficult to make close to
the solar limb has implications not only for studies at high
latitudes, but also for probing deep into the interior. Indeed,
acoustic ray bundles that penetrate deep inside the
Sun connect surface locations that are separated by large
horizontal distances: less signal at high latitudes means
a reduced ability to probe deep at moderate latitudes. In
short, much remains to be learned above 50$^\circ$ heliographic
latitude and for radial distances less than, say, $0.9$~solar radius.

The VIM instrument will return the necessary Doppler images of
the polar regions of the Sun.
In addition, VIM Doppler images combined with Doppler images taken from
Earth will allow to demonstrate the concept of stereoscopic helioseismology.
Thus, VIM data as input for helioseimic methods offer the unique
opportunity to address some outstanding key questions of solar physics:

(1) Internal rotation at high latitudes:
The large-scale differential rotation is an essential dynamical
property of the Sun. Inversions of global-mode
frequency splittings support the idea that rotation is special
at heliographic latitudes above 70$^\circ$. In particular,
some inversions suggest the existence of a buried jet-like
zonal flow (Schou 1999) and a surprisingly slow rotation rate above
75$^\circ$ compared to a smooth extrapolation from lower latitudes
(Schou et al.~1998; Birch \& Kosovichev 1998).
In addition, the near-surface radial
gradient of the angular velocity, may switch sign near
latitude 50$^\circ$ (Corbard \& Thompson 2002).
These results have been derived using
global-mode techniques, which do not necessarily require
local observations at high latitudes. Using observations
out of the ecliptic plane together with techniques of local
helioseismology, VIM may offer the opportunity
to check the validity of the global-mode frequency inversions
at high latitudes. Of course, direct Doppler measurements
will also give important information about the
surface rotation rate.

(2) Meridional circulation:
Surface Doppler measurements have shown the existence
of a meridional flow from the equator to the poles with a
maximum amplitude of about 15 m/s near 30$^\circ$ latitude.
Meridional circulation plays a role in the evolution of
the large-scale surface magnetic field. In flux-transport
dynamo models based on the Babcock-Leighton mechanism,
it carries the poloidal field generated at the surface
down to the bottom of the convection zone, transports
the toroidal magnetic field equatorward, and sets the period
of the solar cycle (Dikpati \& Charbonneau 1999).
Due to mass conservation, the
meridional flow amplitude at the base of the convection
zone is expected to be extremely small.
Techniques of local helioseismology are sensitive to the
first order effect of meridional circulation.
However, the meridional flow at latitudes
above 50$^\circ$ is poorly known and should
be an important target for VIM. One question
is whether the meridional flow has a single-cell or multi-cell
structure. The determination of the latitudinal component
of the meridional flow in the near polar regions is
likely to be extremely difficult (a very small signal is expected).

(3) Polar region variability:
The polar regions appear to be very dynamical. There is
evidence that the differential rotation near the poles may
vary on a short time scale (Ye \& Livingston 1998).
The magnetic properties of
the polar regions at solar minimum determine the strength
of the forthcoming solar maximum (Schatten 1998);
a new sunspot cycle starts in the polar regions.
Thus, studying the dynamics
and variability of the polar regions will help us
understand how the solar dynamo works.
Using global-mode helioseismology,
it was found that there may be a significant
variation of the subsurface high-latitude rotation
as a function of the phase of the solar cycle
(Birch \& Kosovichev 1998; Schou 2003). Does the meridional
circulation, which is known to vary with time near active
latitudes, also vary at high latitudes? In particular, does
it exhibit multi-cell structures that come and go with the
phase of the solar cycle? We may also ask about flow
patterns across the poles. Answering these questions will
require long-term monitoring of the polar regions.

(4) Convection at high latitudes:
Near-surface convective flows act as an important driver
of solar activity. By controlling the evolution of the magnetic
network, supergranulation affects chromospheric
and coronal activity. Near-surface convective flows could
support local dynamo action too. Is the dynamics of supergranulation
different in the polar regions where the effect
of the Coriolis force is expected to be much stronger
than that at lower latitudes? What is the dynamical coupling
between near-surface flows in the near polar regions and
photospheric/coronal magnetism?
Flows at supergranular scales can be measured
with seismic holography or time-distance helioseismology
with sufficient temporal resolution. However, a good spatial
resolution does require that f-modes be observed
(Duvall \& Gizon 2000).
Possibly, flows at larger spatial scales than supergranulation
can be measured with the technique of ring diagram analysis.

(5) Structure and variability of the Tachocline, dynamo flows:
The solar tachocline
is believed to be the seat of the generation of the global
magnetic field (Gilman 2005) and is thus
an important target for helioseismological studies.
A proper determination of the figure
of the solar tachocline obviously requires high latitude
information. It has recently been suggested that prograde
fluid jets may be generated
in the solar tachocline at high latitudes, which may be
measurable by methods of helioseismology (Rempel \& Dikpati 2003;
Christensen-Daalsgard et al.~2004). The tachocline also
appears to support quasi-periodic oscillations with a period
of approximately 1.3 year (Howe et al.~2000), which may be a deep
extension of the torsional oscillations.
There are indications that surface magnetic fields often
cluster at a few specific longitudes for several rotation
periods (Henney \& Harvey 2002).
This may indicate the presence of
an underlying non-axisymmetric magnetic structure and
could be a manifestation of non-axisymmetric modes of
the dynamo (Ruzmaikin 1998; Berdyugina et al.~2006).
Local helioseismology is the appropriate tool to search
for longitudinal variations deep in the convection zone.
Stereoscopic local helioseismology may be an important advance to detect
the signature of deep dynamo flows and their longitudinal
variations inside the Sun.

(6) Stereoscopic helioseismology:
Stereoscopic helioseismology would benefit both global
and local methods of helioseismology.
Stereoscopy requires several observatories.
Both SDO-HMI and GONG (or another ground-based facility)
are expected to be operational at the time of Solar
Orbiter operations. Combining Solar Orbiter with any
one of them will open new windows into the Sun.
Looking at the Sun from two distinct viewing angles results
in an increase of the observed fraction of the Sun's
surface. Global helioseismology would naturally benefit
from this situation, since the modes of oscillation would
be easier to disentangle (reduction of spatial leaks). The
precision on the determination of the mode frequencies
would improve at high frequencies for all spherical harmonic
degrees. Spatial masks for extracting
acoustic modes will be closer to optimal.
With stereoscopic local helioseismology, new acoustic
ray paths can be considered to probe deeper layers in the
interior. Observations from two widely different viewing
angles allow the probing of the solar interior at any
depth, and in particular the solar tachocline (see above). It even
becomes possible to target regions in the solar core.
This aspect of seismic stereoscopy is revolutionary.
The other aspect of stereoscopy is that it is possible to
observe a common area from two different viewing angles.
This case is useful to understand systematic errors
and line-of-sight projection effects in local helioseismology.
It is also potentially important to minimize the contribution
of convection noise in spectra of solar oscillations.
Finally, there is the possibility of observing
the same area with two different spectral lines formed at
two heights in the solar atmosphere to study vertical wave
propagation. In essence, VIM observations in combination with other
solar Doppler imaging have the capability to demonstrate the promising concept
of stereoscopic helioseismology.

\section{Requirements for helioseismology}
Many of the above scientific objectives are difficult to accomplish.
Which objective can be fulfilled and to which extent depends on the
actual performance of the VIM instrument. High quality image stabilization,
accurate knowledge of pointing and precise timing are a
prerequisite for useful measurements.

The observation duration is one of the most critical parameter:
the larger the observation interval $T$, the smaller the
contribution of random noise to helioseismic measurements.
For nearly all quantities of interest,
the level of random noise due to the stochastic
nature of solar oscillations behaves like $T^{-1/2}$.
The current mission baseline of Solar Orbiter
allows observations to be carried out for
about $T=10$~days during each orbit, near perihelion.
Studies of the solar nearsurface
layers require days to weeks of data. Investigations
of the deep interior typically require durations from
weeks to months. In practice, the longer the data coverage,
the better. It would be very meaningful to at least
consider the possibility of making observations during
the cruise phase and outside perihelion passages, even at
a very low telemetry rate. Line-of-sight vectors other than
Earth-Sun are prefered for stereoscopic helioseismology.

The temporal sampling needs to be at least as short as
1 minute to make possible observations of all solar
waves with significant power; faster sampling with, say, 40~s
may be desirable to detect the highest frequency waves.
The use of high-frequency waves is potentially important
to study magnetic effects and to provide more localized
information.

The field of view must be large to resolve the low-degree
modes that penetrate deep into the interior,
as well as to ensure the study of
large scale solar inhomogeneities. A large field of view
also gives the ability to track local areas on the Sun over
long periods of time to compensate for the spacecraft motion
in frames that co-rotate with the Sun. A full-disk
view is ideal, especially to make the most of stereoscopic
analyses.

The spatial resolution measured on the Sun's surface is
a function of ccd pixel size and center-to-limb distance.
Solar oscillations have significant power at all wavelengths
above, say, 5 Mm (f-mode frequency at 3 mHz).
Typically, a spatial sampling of 1 Mm (measured on the
Sun) is sufficient for nearly all applications of helioseismology.
A higher spatial sampling is not needed, while
a lower one is still acceptable for a large range of applications
that involve long-wavelength deep-penetrating
modes of oscillation.

The telemetry rate will likewise have critical impact.
The limited telemetry rates require powerful lossless and lossy data
compression techniques. Intelligent methods of spatial binning,
Fourier filtering and low-pass filtering have to be applied to match the data
volume to the telemetry rates.

\end{document}